\documentclass[aps,twocolumn]{revtex4-1}
\usepackage{amsmath,amssymb}
\usepackage{graphicx}
\usepackage[utf8]{inputenc}
\usepackage{tikz}
\usetikzlibrary{shapes.geometric, decorations.pathmorphing, decorations.markings, arrows.meta}
\usepackage{xcolor}
\usepackage[top=1.0in,bottom=1.0in,left=1.0in,right=1.0in]{geometry}
\usepackage[colorlinks,linkcolor=blue,citecolor=blue]{hyperref}
\usepackage{subfigure}
\usepackage{float}
\usepackage{placeins}
\usepackage{fancyhdr}

\newcommand{\be}{\begin{equation}}
\newcommand{\ee}{\end{equation}}
\newcommand{\bea}{\begin{eqnarray}}
\newcommand{\eea}{\end{eqnarray}}
\newcommand{\msbar}{{\overline{\mbox{\rm MS}}}}

\begin{document}

%\title{Deeply virtual Compton scattering  in holographic QCD: \\
%Quark and Gluon GPDs at non-zero skewness}

%\title{GPDs from holographic moment parametrization at finite skewness: \\a comparison to lattice QCD}

\title{Net-Proton Number Cumulants in Strongly and Weakly Coupled QGP
}

\author{Kiminad A. Mamo}
\email{kamamo@wm.edu}
\affiliation{
Physics Department, William \& Mary, Williamsburg, VA 23187, USA}

\date{\today}

\begin{abstract}

We provide a description of the QCD phase diagram across different energy regimes. Without any adjustable free parameters fitted to lattice or experimental data, we demonstrate that the net-proton number cumulant ratios from the Beam Energy Scan-I (BES-I) central ($0-5\%$) Au+Au collision experiments at RHIC, for center-of-mass energies $\sqrt{s} \geq 39\,\text{GeV}$, are well described by the $\textit{dynamical soft-wall}$ AdS black hole model, indicating the formation of a strongly coupled quark-gluon plasma (sQGP) dominated by gluons in the large $N_c$ limit. We identify a critical point at $\mu_B = 247\,\text{MeV}$ or $\sqrt{s}=16\,\text{GeV}$ in this model, where net-proton cumulant ratios show non-monotonic variation. However, for $\sqrt{s} < 39\,\text{GeV}$, the recent BES-II data does not exhibit the expected variation due to the critical point, suggesting instead a transition to a nearly-cold ($\mu_B \gg T$) weakly coupled QGP (wQGP), for $\sqrt{s} < 7.7\,\text{GeV}$, dominated by valence quarks. This prediction can be tested by the upcoming STAR Fixed-Target experimental data.

\end{abstract}

\maketitle

\section{Introduction}
In high-energy nuclear collisions \cite{Gyulassy:2004zy, BraunMunzinger:2007zz, Jacak:2012dx}, quarks can become deconfined, forming a Quark-Gluon Plasma (QGP) \cite{Itoh:1970uw, Collins:1974ky, Cabibbo:1975ig, Chapline:1976gy, Shuryak:1978ij}, a state believed to have existed in the early Universe \cite{Boyanovsky:2006bf}. Evidence for QGP comes from heavy-ion collision experiments at facilities such as the CERN SPS \cite{Heinz:2000bk}, BNL RHIC \cite{Adams:2005dq, Arsene:2004fa, Adcox:2004mh, Back:2004je, Gyulassy:2004zy, Jacak:2012dx}, and CERN LHC \cite{Muller:2012zq, Schukraft:2013wba, Braun-Munzinger:2015hba}. These experiments revealed that QGP behaves as a strongly interacting fluid, termed sQGP \cite{Shuryak:2003ty, Shuryak:2014zxa}. 

The QCD phase diagram features sQGP at high temperatures and hadronic matter at lower temperatures \cite{Fukushima:2010bq, BraunMunzinger:2009zz, Asakawa:1989bq}. Lattice QCD calculations suggest a crossover transition at low baryon chemical potential ($\mu_{\mathrm B}<300\,\text{MeV}$) and a pseudo-critical temperature ($T_{pc}\sim 155 \,\text{MeV}$) \cite{Aoki:2006we, Aoki:2009sc, Bazavov:2011nk, Gupta:2011wh, HotQCD:2018pds}. Similar to phase diagrams in condensed-matter systems, the QCD phase diagram can be explored by varying the baryon chemical potential ($\mu_{\mathrm B}$) through changes in heavy-ion collision energy \cite{Bzdak:2019pkr, Luo:2017faz}. This is the goal of the Beam Energy Scan (BES) and STAR Fixed-Target programs at RHIC \cite{Odyniec:2013aaa, STAR:2020tga, Pandav:2024cpod, Meehan:2016qon}. 

%%%%%%%%%%%%%%%%%%%%%55

Since hadron yields and multiplicity distributions related to conserved quantities in high-energy nuclear collisions indicate thermodynamic equilibrium \cite{Cleymans:2005xv, Andronic:2017pug, Braun-Munzinger:2011shf, Bazavov:2012vg, Borsanyi:2014ewa, Adamczyk:2017iwn, Andronic:2017pug, Gupta:2020pjd}, a critical point in the QCD phase diagram is expected to cause a divergence in the correlation length. This leads to the divergence of higher-order cumulants, which scale with higher powers of the correlation length \cite{Stephanov:2008qz}. Additionally, near the critical point, non-Gaussian net-baryon number distributions and divergent susceptibilities are anticipated, resulting in moments that exhibit non-monotonic variation as a function of $\sqrt{s}$ \cite{Stephanov:1999zu, Hatta:2002sj, Stephanov:2008qz, Stokic:2008jh, Asakawa:2009aj, Stephanov:2011pb, Gavai:2010zn}. Various models have previously addressed the location of the critical point in QCD including \cite{Xin:2014ela,Gao:2016qkh,Qin:2010nq,Shi:2014zpa,Fischer:2014ata,Gao:2020qsj,Asakawa:1989bq,Schwarz:1999dj,Li:2018ygx,Zhuang:2000ub,Fu:2019hdw,Zhang:2017icm,Fu:2021oaw,Hippert:2023bel,Cai:2022omk,Li:2023mpv}.

In this paper, using the holographic approach~\cite{Maldacena:1997re, Casalderrey-Solana:2011dxg, DeWolfe:2013cua,Karch:2006pv,Li:2013oda,Ballon-Bayona:2020xls}, see \cite{Natsuume:2014sfa} for a review, we identify a critical point in the sQGP at \(\mu_B = 247\,\text{MeV}\) or \(\sqrt{s}=16\,\text{GeV}\) on the freeze-out line. We demonstrate that the net-proton number cumulant ratios exhibit non-monotonic variation at this point. Our comparison with the combined BES-I and BES-II experimental data~\cite{Pandav:2024cpod} shows that the sQGP description is accurate for $\sqrt{s} \geq 39\, \text{GeV}$, confirming the formation of sQGP in this high-energy regime which is dominated by gluons in the large $N_c$ limit. However, the sQGP description breaks down for $\sqrt{s} < 39\, \text{GeV}$, as the experimental data does not exhibit the expected non-monotonic variation. Based on this observation, we conjecture the formation of a nearly-cold ($\mu_B \gg T$) weakly coupled QGP (wQGP) in the low-energy regime ($\sqrt{s} < 7.7\,\text{GeV}$) dominated by valence quarks. This conjecture and its predictions can be tested by the upcoming data from the STAR Fixed-Target experiments at RHIC~\cite{Meehan:2016qon}.

\section{Strongly Coupled QGP (sQGP)} 
Our theoretical framework for studying strongly coupled QGP (sQGP) utilizes the \textit{dynamical soft-wall} AdS black hole model \cite{Ballon-Bayona:2020xls}, which provides a holographic dual to sQGP based on the original \textit{soft-wall} model \cite{Karch:2006pv}. The Einstein-Maxwell-Dilaton (EMD) action in the Einstein frame is given by \cite{Ballon-Bayona:2020xls}:

\bea \label{EMD action}
S &=& \dfrac{1}{16\pi G_5}\int d^{5}x \sqrt{-g}\left(R - \dfrac{4}{3}g^{\mu\nu}\partial_{\mu}\phi\partial_{\nu}\phi + V(\phi)\right)\nonumber\\
&-&\dfrac{1}{4g_5^2}\int d^{5}x \sqrt{-g}F_{\mu\nu}F^{\mu\nu} ,
\eea
where \(G_5 \sim 1/N_c^2\) is the 5-dimensional Newton's constant, \(g_5^2 \sim 1/N_c N_f\) is the bulk flavor gauge coupling constant (and to match the action in \cite{Ballon-Bayona:2020xls} we will assume $16\pi G_5/g_5^2\propto\frac{N_f}{N_c}\approx 1$ for $N_c=N_f=3$), \(g=\mathrm{det}(g_{\mu\nu})\), \(R\) is the Ricci scalar, \(\phi\) is the dilaton field, \(V(\phi)\) is the dilaton potential, and \(F_{\mu \nu}\) is the electromagnetic field strength defined by \(F_{\mu \nu} = \partial_{\mu} A_{\nu} - \partial_{\nu} A_{\mu}\). For small \(\phi\), the dilaton potential is \(V(\phi)= 12 - (4/3) m^2 \phi^2\), with \(m^2 = -4\). The full dilaton potential, at zero temperature and chemical potential, is detailed in section 4 (Model-IA) of \cite{Li:2013oda}. The \textit{dynamical soft-wall} model generalizes the original \textit{soft-wall} model \cite{Karch:2006pv} by dynamically including the back-reaction of the wall (formed by the gluon condensate) on the geometry without introducing any new parameter other than the mass-scale $\kappa$, which is fixed to be $\kappa=0.430\,\text{GeV}$ based on the Regge slope of meson spectra in Model-IA of \cite{Li:2013oda}. Notably, this value of $\kappa$ is very close to the experimentally extracted value of $\kappa=0.402\,\text{GeV}$ in \cite{Mamo:2021jhj} within the original \textit{soft-wall} model framework. Here, we also work in the limit $\kappa \gg \sigma^{1/3}, m_q$, allowing us to ignore the chiral symmetry breaking effects, since the chiral condensate $\sigma^{1/3} = 0.180\,\text{GeV}$, and mass of the quarks $m_q=0.006\,\text{GeV}$ in the \textit{dynamical soft-wall} model (see Model-IA of \cite{Li:2013oda}).

Using the AdS/CFT correspondence dictionary \cite{Witten:1998qj, Gubser:1998bc}, which relates \(m^2 = \Delta (\Delta - 4)\), we find \(\Delta = 2\) for \(m^2 = -4\), identified as the dimension of the gluonic operator \({\cal O} = \text{tr}\,F_{YM}^2\) in pure Yang-Mills theory in the IR which is strongly coupled. In the UV, \(\mathcal{N}=4\) super Yang-Mills theory (dual to Type IIB supergravity) has \(\Delta=4\) for \({\cal O} = \text{tr}\,F_{YM}^2\). We conjecture that, after supersymmetry breaking in \(\mathcal{N}=4\) super Yang-Mills theory, \(\Delta\) flows to \(\Delta=2\) in the IR, where the strongly coupled pure Yang-Mills theory is its low-energy effective field theory. The exact mechanism of supersymmetry breaking is unknown, but we postulate that it generates the dilaton potential in this model, whose predictions we study. Note that the \textit{dynamical soft-wall} model is not the only holographic model featuring a potential for the dilaton with \(\Delta<4\) for \({\cal O} = \text{tr}\,F_{YM}^2\); see, for example, \cite{Gubser:2008ny, Gubser:2008yx, Gubser:2008sz, DeWolfe:2010he, Rougemont:2023gfz}.

The \textit{dynamical soft-wall} AdS black hole solution to the field equations derived from the action \eqref{EMD action}, which provides a holographic dual to the strongly coupled QGP (sQGP), is given by \cite{Ballon-Bayona:2020xls}:
\begin{eqnarray}
ds^2 &=& \dfrac{1}{\zeta(z)^2}\left(\dfrac{dz^2}{f(z)} - f(z)dt^2 + d\vec{x}^2\right), \label{Metriczeta} \\
\phi(z) &=& \kappa^2 z^2 , \label{phi} \\ 
A_t(z) &=& \mu \left( 1 - \frac{C_2(z)}{C_2(z_h)} \right), \label{At}
\end{eqnarray}
where, for $\kappa^2 z^2 \ll 1$, we have the analytic results \cite{Ballon-Bayona:2020xls}:
\begin{eqnarray}
\zeta(z) &=& z \left( 1  + \frac{4}{45} (\kappa^2 z^2)^2 + {\cal O} (\kappa^2 z^2)^4 \right), \label{zetaExp} \\
f(z) &=& 1 - \dfrac{C_4(z)}{C_4(z_h)} \nonumber \\
&+& \left(\frac{16\pi G_5}{g_5^2}\right)\dfrac{\mu^2}{C_4(z_h) C_2(z_h)^2} \Big[ C_4(z_h) C_6(z)\nonumber\\
&-& C_4(z) C_6(z_h) \Big], \label{horfun}
\end{eqnarray}
with $z=z_h$ as the location of the AdS black hole horizon, for $\kappa^2 z^2 \ll 1$,
\begin{eqnarray}
C_2(z) &=& \frac{z^2}{2} \left( 1 + \frac{4}{135} (\kappa^2 z^2)^2 + {\cal O} (\kappa^2 z^2)^4 \right), \nonumber \\
C_4(z) &=& \frac{z^4}{4} \left( 1 + \frac{2}{15} (\kappa^2 z^2)^2 + {\cal O} (\kappa^2 z^2)^4 \right), \nonumber \\
C_6(z) &=& \frac{z^6}{12} \left( 1 + \frac{8}{45} (\kappa^2 z^2)^2 + {\cal O} (\kappa^2 z^2)^4 \right)\,.
\end{eqnarray}

%%%%%%%%%%%%%%
Throughout this paper, we define the baryonic chemical potential as $\mu_B \equiv \mu$. Consequently, the bulk gauge field $A_{\mu} \equiv A_{\mu}^{B}$ transforms under the $U_{B}(1) \subset U(N_f)$ subgroup, which corresponds to baryon number conservation. The baryon charge density (or baryon number density) is then defined as
\be
n_B(T, \mu_B) = \frac{1}{N_c} n(T, \mu_B),
\ee
where $n(T, \mu_B)$ represents the total valence quark charge density (or total valence quark number density) at finite baryon chemical potential $\mu = \mu_B$. At strong coupling, this quantity is given by
\bea
n(T, \mu_B) &=& \langle j^t \rangle = \frac{\delta S}{\delta A_t} = - \frac{1}{g_5^2} \left[ \frac{1}{\zeta(z)} \partial_z A_t \right]_{z=\epsilon}\,,\nonumber\\
&=& \frac{1}{g_5^2} \frac{\mu_B}{C_2(z_h)}\,,
\eea
where the bulk flavor gauge coupling is $\frac{1}{g_5^2} \propto N_c N_f$. At the boundary $z = 0$, we assume the potential satisfies $A_t^{B}(z=0) = A_t(z=0) = \mu_B$, which couples to the quark charge density operator $j^t \equiv \sum_{q=1}^{N_f} \bar{q} \gamma^t q$ through the interaction term
\be\label{INT}
S_{QCD}^{\text{int}} = A_t^{B}(0) j^t = \mu_B \sum_{q=1}^{N_f} \bar{q} \gamma^t q.
\ee
This interaction term \eqref{INT} establishes $\mu_B$ as a background baryon chemical potential sourced by baryons (three valence quarks collectively) rather than by individual valence quarks. Therefore, the interaction term \eqref{INT} aligns with the lattice interaction term (as used, for instance, in \cite{HotQCD:2018pds} and \cite{Borsanyi:2020fev}), with the substitution 
\be\label{replace}
\mu_B \rightarrow \frac{1}{3} \mu_B^{\text{lattice}}.
\ee

This substitution is necessary since lattice QCD works, for example \cite{HotQCD:2018pds, Borsanyi:2020fev}, assume that the background chemical potential is sourced by individual valence quarks. Notably, with our convention \eqref{INT}, the critical point in the fRG approach \cite{Fu:2021oaw} shifts from $\mu_c = 635~\text{MeV}$ to $\mu_c = \frac{1}{3} \times 635~\text{MeV} = 212~\text{MeV}$, aligning closely with the critical point in our \textit{dynamical soft-wall} AdS black hole model ($\mu_c = 247~\text{MeV}$). Similarly, using this convention in black hole engineering models \cite{Hippert:2023bel, Cai:2022omk, Li:2023mpv} reduces the critical point from $\mu_c \approx 589~\text{MeV}$ to $\mu_c \approx \frac{1}{3} \times 589~\text{MeV} = 196~\text{MeV}$, as these models are calibrated to lattice data, which requires the substitution \eqref{replace}. The same conclusion applies to other models (including holographic ones) fitted to lattice data at small baryon chemical potential.

%%%%%%%%%%%%%%%%5

The Hawking temperature \( T \) and entropy density \( s \) of the \textit{dynamical soft-wall} AdS black hole, for $\kappa^2 z^2 \ll 1$, are
\begin{align}
\tilde{T}(k_h, \tilde{\mu}) &= \frac{1}{\sqrt{k_h} \pi}\Biggr(1 + \frac{2 k_h^2}{15} -\Big[\frac{k_h \tilde{\mu}^2}{6} + \frac{14 k_h^3 \tilde{\mu}^2}{405}\nonumber\\
&+ \frac{2 k_h^5 \tilde{\mu}^2}{1215}\Big]\left(\frac{16\pi G_5}{g_5^2}\right)\Bigg)\,, \label{TH} \\
\tilde{s}(k_h) &= \frac{1}{k_h^{3/2}} \left( 1 - \frac{4 k_h^2}{15} \right)\,, \label{SH}
\end{align}
where the dimensionless combinations are defined as \(\tilde{T} \equiv T/\kappa\), \(\tilde{s} \equiv 4G_5 \times s/\kappa^3\), \(\tilde{\mu} \equiv \mu/\kappa\), and \(k_h \equiv \kappa^2 z_h^2\). The Hawking temperature has a local minimum \(\tilde{T}_{\min}(\tilde{\mu}) = \tilde{T}(k_{h \min}(\tilde{\mu}), \tilde{\mu})\), defined by \(\partial_{k_h} \tilde{T}(k_h, \tilde{\mu}) = 0\) at \(k_h = k_{h \min}(\tilde{\mu})\). 

\begin{figure}
%[ht!]
\centering
\subfigure[\label{kh-vs-T}]{%
\includegraphics[height=4cm,width=8.1cm]{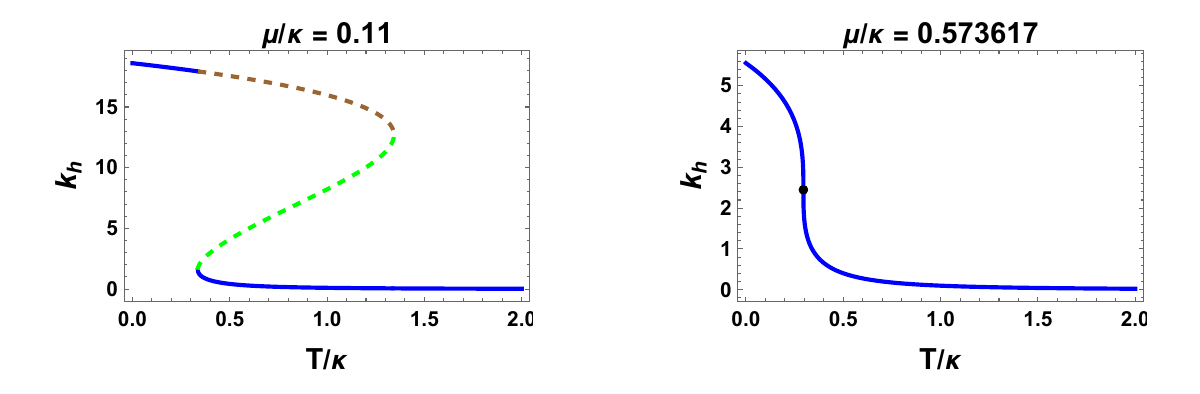}%
}
\hfill
\subfigure[\label{f-vs-T}]{%
\includegraphics[height=4cm,width=8.1cm]{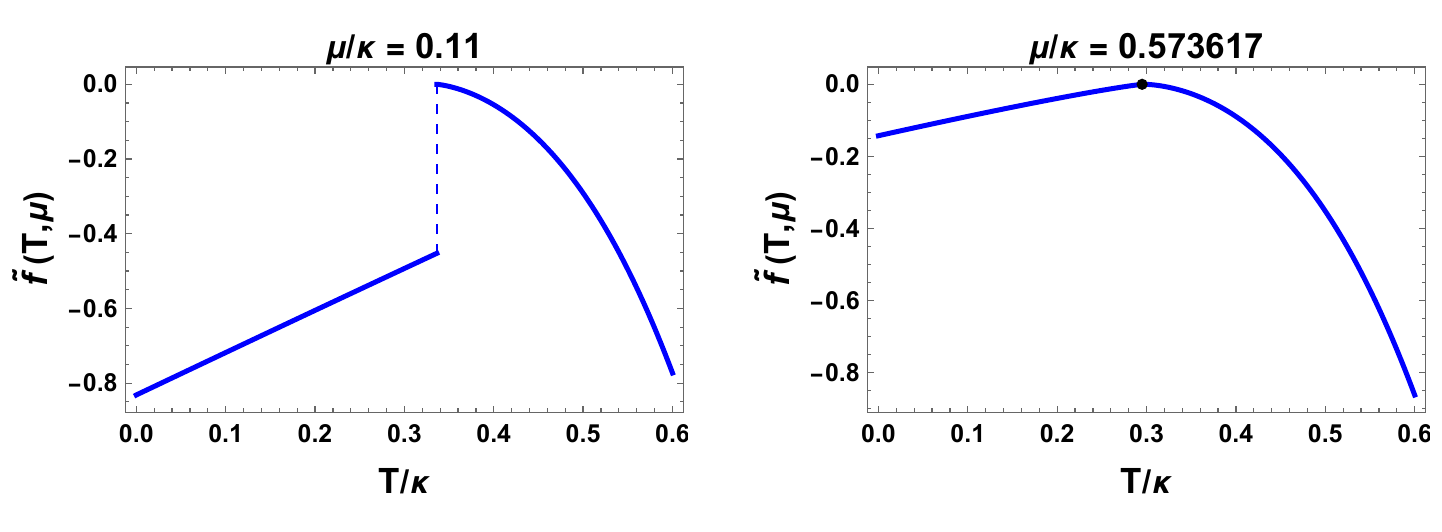}%
}
\caption{(a) The location of the \textit{dynamical soft-wall} AdS black hole horizon $k_h=\kappa^2z_h^2$ versus the Hawking temperature $T/\kappa$, given in (\ref{TH}), at different chemical potentials $\mu/\kappa$. (b) The corresponding free energy density $\tilde{f}$, given in (\ref{FreeEnergyBH}), versus $T/\kappa$. In both (a) and (b), the \textcolor{blue}{blue} curves above $T_{min}/\kappa$ corresponds to the large AdS black hole branches, while the \textcolor{blue}{blue} curves below $T_{min}/\kappa$ are the physical small AdS black hole branches. They are smoothly connected to each other at the critical point $(T_c/\kappa,\mu_c/\kappa)=(0.295055,0.573617)$ denoted by the black dot where we have a second-order phase transition between them. The critical point also represents the end-point of the first-order phase transition from the large to small AdS black hole. The \textcolor{brown}{brown} and \textcolor{green}{green} dashed curves, in (a), are the nonphysical small AdS black hole branches.}
\label{HUD}
\end{figure}

In Fig.\,\ref{kh-vs-T}, we have displayed the three roots or branches \(k_h(T, \mu, n)\), where the \textcolor{blue}{Blue}, \textcolor{green}{Green}, and \textcolor{brown}{Brown} curves are the first ($n=1$), second ($n=2$), and third ($n=3$) roots, respectively, determined by inverting (\ref{TH}) for $\frac{16\pi G_5}{g_5^2}\approx 1$.

%\begin{widetext}

%\begin{widetext}
\begin{figure}
%[ht!]
\centering
\subfigure[\label{BH-phases}]{%
\includegraphics[height=4.5cm,width=7cm]{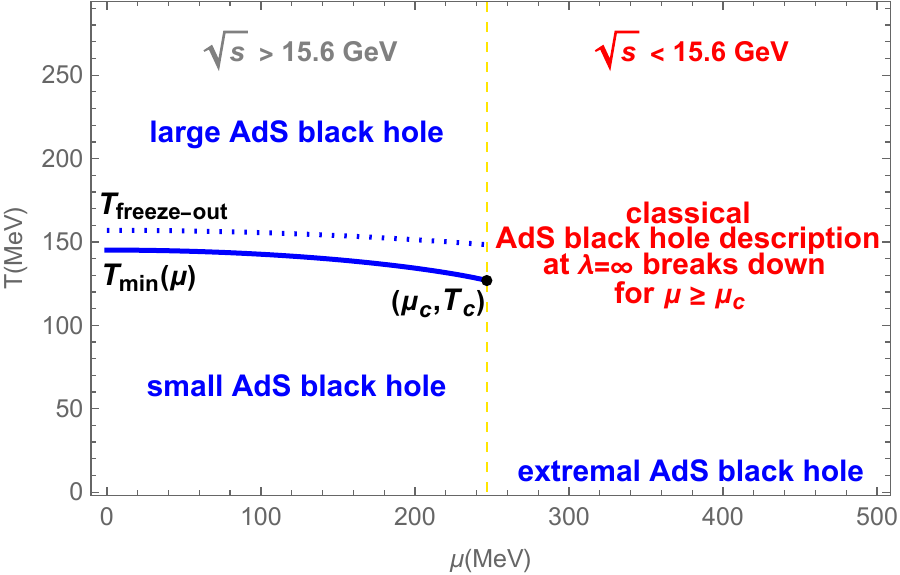}%
}
\caption{The phase diagram of the \textit{dynamical soft-wall} AdS black hole with $\kappa=0.430\,\text{GeV}$ fixed by the Regge slope of meson spectra \cite{Li:2013oda}. The classical (large-$N_c$) AdS black hole description breaks down for $\mu >\mu_c$, we need quantum (finite $1/N_c$) correction in this regime. Note that, according to the experimental data shown in Fig.\,\ref{CQCP}, the large AdS black hole does not undergo a first-order phase transition to the physical small AdS black hole. Instead, the large AdS black hole freezes out or evaporates at $T_{freeze-out} \approx 157\,\text{MeV} > T_{min} \approx 145\,\text{MeV}$ by emitting Hawking radiation (hadron resonances). Therefore, according to the experimental data, we conclude that the physical small AdS black hole (the AdS black hole solution for $T<T_{min}$) has lower pressure density (or higher free energy density) than the thermal gas (Hawking radiation) in AdS (which is the holographic dual to hadron resonance gas).}
\label{BH-phases}
\end{figure}

%\begin{widetext}
\begin{figure}
%[ht!]
\centering
\subfigure[\label{CQCP}]{%
\includegraphics[height=12.5cm,width=7.5cm]{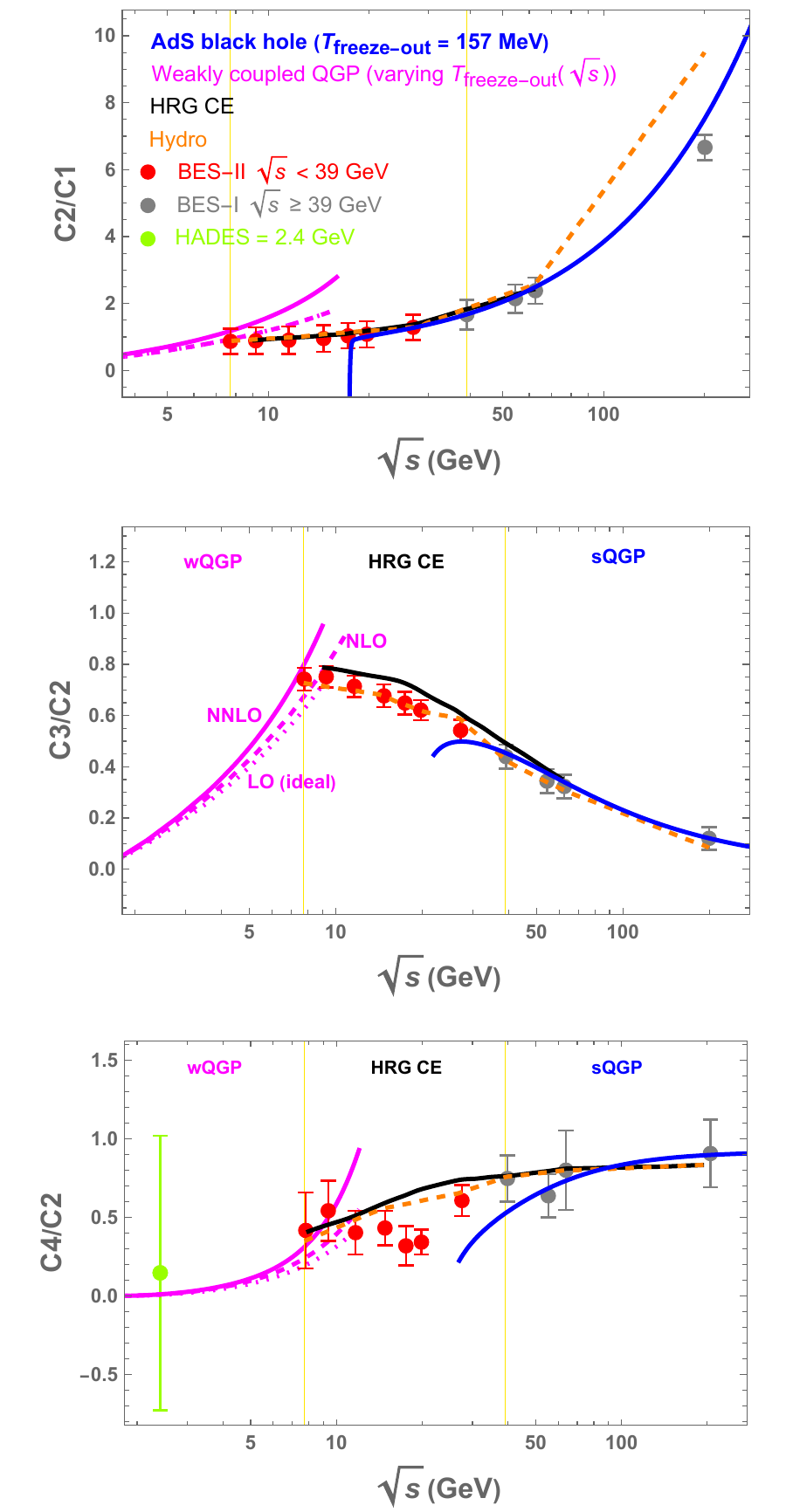}%
}
\caption{Net-proton number cumulant ratios as defined in (\ref{firstC}), and comparison to experimental data from BES-I (\textcolor{gray}{gray} data points) \cite{STAR:2020tga}, and BES-II (\textcolor{red}{red} data points) \cite{Pandav:2024cpod} for $0-5\%$ Au+Au collisions at $7.7\,\text{GeV}\leq\sqrt{s}\leq200\,\text{GeV}$, and our predictions for \textcolor{magenta}{Fixed-Target} BES experiments at $\sqrt{s}<7.7\,\text{GeV}$. The \textcolor{green}{green} data point is from HADES \cite{HADES:2020wpc}. The \textcolor{orange}{orange} dashed, and \textcolor{black}{black} curves are the hydrodynamics (\textcolor{orange}{Hydro}) \cite{Vovchenko:2021kxx}, and hadron resonance gas (HRG) \cite{Braun-Munzinger:2020jbk} results, respectively, with fixed or conserved baryon charge in the canonical ensemble (CE). The \textcolor{blue}{blue}, and \textcolor{magenta}{magenta} curves are based on our strongly (\ref{FreeEnergyBH}) and weakly (\ref{masslessp}) coupled results, respectively. See text.}
\label{CQCP}
\end{figure}

The free energy density $f$ or pressure density $p=-f$, is given by integrating the first-law of thermodynamics (in the grand canonical ensemble) at fixed $\mu$, and volume $V$ as
\begin{align}\label{FreeEnergyBH0}
\tilde{f}(T, \mu) &= -\int_{\tilde{T}_{min}}^{\tilde{T}} \tilde{s}(\tilde{T}, \tilde{\mu}) \, d\tilde{T}'\,,\nonumber\\
&=-\int_{k_{hmin}(\mu)}^{k_h(T,\mu)} \tilde{s}(k_h') \left( \frac{\partial \tilde{T}(k_h', \tilde{\mu})}{\partial k_h'} \right) dk_h'\,,\nonumber\\
\end{align}
where we have defined the dimensionless combination $\tilde{f}=4G_5\times f/\kappa^4$, and we have fixed the integration constant in such a way that the free energy density vanishes at $\tilde{T}=\tilde{T}_{min}=\tilde{T}(k_{hmin},\tilde\mu)$. More explicitly, the free energy density is given by:
\bea\label{FreeEnergyBH}
\tilde{f}(T, \mu) &=& -\frac{1}{k_h^4} \Bigg[ \frac{k_h^2}{4 \pi} \left( 1 - \frac{k_h^2}{k_{hmin}^2} \right) \nonumber \\
&+& \frac{1}{3\pi}k_h^4 \log \left( \frac{k_h}{k_{hmin}} \right)\nonumber\\
&-& \frac{2 }{75 \pi} k_h^6\left( 1-\frac{k_{hmin}^2}{k_h^2}\right) \nonumber \\
&+& \Big[\frac{k_h^3 \tilde{\mu}^2}{12 \pi} \left( 1 - \frac{k_h}{k_{hmin}} \right) \nonumber \\
&-& \frac{26 }{405 \pi} k_h^5 \tilde{\mu}^2\left( 1-\frac{k_{hmin}}{k_h} \right) \nonumber \\
&+& \frac{19}{3645 \pi}k_h^7 \tilde{\mu}^2\left( 1 - \frac{k_{hmin}^3}{k_h^3} \right) \nonumber \\
&+& \frac{4 }{10125 \pi} k_h^9 \tilde{\mu}^2\left( 1 - \frac{k_{hmin}^5}{k_h^5} \right)\Big]\left(\frac{16\pi G_5}{g_5^2}\right) \Bigg]\,,\nonumber\\
\eea
where we set $\frac{16\pi G_5}{g_5^2}\approx 1$ for $N_c=N_f=3$, and, in Mathematica, $k_{h}=k_{h}(\tilde{T},\tilde{\mu},n=1)$ is given by the Root function
\bea
&&k_{h}(\text{$\tilde{T}$},\text{$\tilde{\mu}$},\text{n})\text{:=}\text{Root}\Big[16 \text{$\#$1}^{10} \tilde{\mu}^4+672 \text{$\#$1}^8 \tilde{\mu}^4\nonumber\\
&-&2592 \text{$\#$1}^7 \tilde{\mu}^2+10296 \text{$\#$1}^6 \tilde{\mu}^4 -73872 \text{$\#$1}^5 \tilde{\mu}^2\nonumber\\
&+&\text{$\#$1}^4 \left(68040 \tilde{\mu}^4+104976\right)-670680 \text{$\#$1}^3 \tilde{\mu}^2\nonumber\\
&+& \text{$\#$1}^2 \left(164025 \tilde{\mu}^4+1574640\right)\nonumber\\
&+&\text{$\#$1} \left(-1968300 \tilde{\mu}^2-5904900 \pi ^2 \tilde{T}^2\right)+5904900\&,n\Big]\,,\nonumber\\
\eea
for $n=1,2,3$, and \(k_{hmin}=k_{hmin}(\tilde{\mu})\) is given by the Root function
\bea
k_{hmin}(\tilde{\mu})&\text{:=}&\text{Root}\Big[12 \text{$\#$1}^5 \tilde{\mu}^2+140 \text{$\#$1}^3 \tilde{\mu}^2-324 \text{$\#$1}^2\nonumber\\
&+&135 \text{$\#$1} \tilde{\mu}^2+810\&,2\Big]\,.
\eea
We have plotted the resulting free energy density $\tilde{f}(T,\mu)$ in Fig.\,\ref{f-vs-T}, using the first root $n=1$ (\textcolor{blue}{Blue} curve). The phase diagram of the \textit{dynamical soft-wall} AdS black hole, with $\kappa=0.430\,\text{GeV}$ fixed by the Regge slope of meson spectra \cite{Li:2013oda}, is shown in Fig.\,\ref{BH-phases}.

\section{Weakly Coupled QGP (wQGP)}
The weakly coupled quark-gluon plasma (wQGP), in the nearly-cold $\mu_B=\mu\gg T$ limit (based on our convention of the interaction term \eqref{INT}), can be characterized by the free energy density approximately given by \cite{Kurkela:2009gj}
\begin{align}
f(T,\mu) &\approx -\frac{N_c N_f}{4 \pi^2}\mu^4\Bigg\{\frac{1}{3} - \frac{d_A}{N_c}\frac{\alpha_s(\bar{\Lambda})}{4\pi} \nonumber \\
&+ \frac{d_A}{N_c}\bigg[\frac{4}{3}\left(N_f-\frac{11C_A}{2}\right)\ln\frac{\bar{\Lambda}}{2\mu}\nonumber\\
&- \frac{142}{9}C_A + \frac{17}{2}C_F + \frac{22}{9}N_f\bigg]\frac{\alpha_s^2(\bar{\Lambda})}{(4\pi)^2}\Bigg\}\,,\nonumber\\
&\approx  -\frac{N_c N_f}{4 \pi^2}\mu^4\Bigg\{\frac{1}{3} - \frac{d_A}{N_c^2}\frac{\lambda_s(\bar{\Lambda})}{(4\pi)^2} \nonumber \\
& + \frac{d_A}{N_c^3}\bigg[\frac{4}{3}\left(N_f-\frac{11C_A}{2}\right)\ln\frac{\bar{\Lambda}}{2\mu}\nonumber\\
&- \frac{142}{9}C_A + \frac{17}{2}C_F + \frac{22}{9}N_f\bigg]\frac{\lambda_s^2(\bar{\Lambda})}{(4\pi)^4}\Bigg\}\,,\nonumber\\
\label{masslessp}
\end{align}
for massless quarks with $N_c = N_f = 3$, $d_A = N_c^2 - 1$, $C_A = N_c$, $C_F = \frac{N_c^2 - 1}{2N_c}$, and the QCD running coupling constant \cite{Kurkela:2009gj}
\begin{align}
\alpha_s(\bar{\Lambda}) &= \frac{4\pi}{\beta_0 \log\left(\bar{\Lambda}^2 / \Lambda_{\tiny \msbar}^2\right)} \nonumber\\
&\times \left( 1 - \frac{2\beta_1}{\beta_0^2} \frac{\log\left(\log\left(\bar{\Lambda}^2 / \Lambda_{\tiny \msbar}^2\right)\right)}{\log\left(\bar{\Lambda}^2 / \Lambda_{\tiny \msbar}^2\right)} \right)\,, 
\end{align}
with $\Lambda_{\tiny \msbar} = 0.378\,\text{GeV}$, $\beta_0 = \frac{11}{3} C_A - \frac{2}{3} N_f$, $\beta_1 = \frac{17}{3} C_A^2 - C_F N_f - \frac{5}{3} C_A N_f$, and the running energy scale 
\be
\bar{\Lambda} \equiv 2\pi\sqrt{T^2 + \frac{\mu^2}{\pi^2}}\,.
\ee
In the last line of \eqref{masslessp}, we have defined the 't Hooft coupling constant of QCD $\lambda_{s}(\bar{\Lambda})\equiv\alpha_s(\bar{\Lambda})\times 4\pi\times N_c$.

Note that the free energy density at weak coupling \eqref{masslessp} is $f\propto N_c N_f$ in the large $N_c$ limit, while the free energy density at strong coupling \eqref{FreeEnergyBH} $f\propto N_c^2$ in the large $N_c$ limit which indicates that at large chemical potential $\mu\gg T$ the free energy density is dominated by valence quarks while at small chemical potential $\mu\ll T$ the free energy density is dominated by gluons. 

\section{Comparison and Predictions for BES and STAR Fixed-Target Experiments}
The net-proton ($N_{p} - N_{\bar{p}} = \Delta N_{p}$) distribution is obtained by measuring the number of protons ($N_{p}$) and anti-protons ($N_{\bar{p}}$) in high-energy nuclear collisions at various $\sqrt{s}$ \cite{Braun-Munzinger:2011shf}. The higher order cumulants ($C_{n}$) of the net-proton distribution that can be measured include: $C_{1} = M$, $C_{2} = \sigma^2$, $C_{3} = S\sigma^{3}$, and $C_{4} = \tilde{\kappa}\sigma^{4}$. The higher order moments are skewness (${\it S}$), defined as ${\it S} = \left\langle (\delta N)^3 \right\rangle/\sigma^{3}$, and kurtosis ($\tilde{\kappa}$), defined as $\tilde{\kappa} = \left[\left\langle (\delta N)^4 \right\rangle/\sigma^{4}\right] - 3$, where $\delta N = N - M$, $M$ is the mean, and $\sigma$ is the standard deviation.

Ratios of cumulants are related to baryon-number susceptibilities, $\chi_{n}$ = -$\frac{d^n f}{d\mu^n}$, where $n$ is the order and $f$ is the system's free energy density at a given $T$ and $\mu$. For example, $C_{3}/C_{2}$ = ${\it{S}}\sigma$ = $(\chi_{3}/T)/(\chi_{2}/T^2)$ and $C_{4}/C_{2}$ = $\tilde{\kappa}\sigma^2$ = $(\chi_{4})/(\chi_{2}/T^2)$. In general, on the freeze-out line, we have
\begin{align}
\frac{C_n}{C_m} &\equiv \frac{T^n}{T^m} \times \left. \frac{\partial_{\mu}^n \left[ -f(T,\mu)/T^4 \right]}{\partial_{\mu}^m \left[ -f(T,\mu)/T^4 \right]} \right|_{T=T_{\text{freeze-out}}} \label{firstC}\,.
\end{align}

In Fig.\,\ref{CQCP}, the \textcolor{blue}{blue} curve shows the net-proton number cumulant ratios of the \textit{dynamical soft-wall} AdS black hole (sQGP)
\begin{align}
\frac{C_n}{C_m}
&= \frac{T^n}{T^m} \times \left. \frac{\partial_{\mu}^n \left[ \tilde{f}(T, \mu) \right]}{\partial_{\mu}^m \left[ \tilde{f}(T, \mu) \right]} \right|_{T=T_{\text{freeze-out}}} \label{secondC}\,,
\end{align}
where $\tilde{f}(T,\mu)$ is the free energy density of the AdS black hole (\ref{FreeEnergyBH}) (Note that, for our numerical implementation of (\ref{secondC}), we have Taylor expanded (\ref{FreeEnergyBH}) upto $\mathcal{O}(\mu^{200})$ terms). The \textcolor{blue}{blue} curves, in In Fig.\,\ref{CQCP}, assume a fixed freeze-out temperature \(T_{\text{freeze-out}} \gg \mu\) with $T_{\text{freeze-out}} = 0.157\,\text{GeV}$, and 
\begin{equation}\label{mufixedTf}
\frac{\mu(\sqrt{s})}{0.157\,\text{GeV}} = \frac{21.354}{1 + 0.794\,\text{GeV}^{-1} \sqrt{s}}\,,
\end{equation}
determined from the fit to the red data points (from first moments) of Figure 13 in \cite{Braun-Munzinger:2020jbk}, see Fig.\,\ref{muTs}. Note that, on the freeze-out line at $T=T_{\text{freeze-out}} = 0.157\,\text{GeV}$, the net-proton number cumulant ratios of the \textit{dynamical soft-wall} AdS black hole in \eqref{secondC} diverge near the critical point $\mu=\mu_c=247\,\text{MeV}$ or $\sqrt{s}=16\,\text{GeV}$. Specifically, on the freeze-out line at $T=T_{\text{freeze-out}} = 0.157\,\text{GeV}$, the baryon-number susceptibility diverges as $\chi_2\propto(\mu-\mu_c)^{-\delta}$ with critical exponent $\delta=1$. Hence, the higher order baryon-number susceptibilities diverge as $\chi_n\propto(\mu-\mu_c)^{-(\delta+(n-2))}$, and the net-proton number cumulant ratios diverge as $C_n/C_m\propto(\mu-\mu_c)^{m-n}$ for $m<n$.
%\begin{widetext}
\begin{figure}
%[ht!]
\centering
\subfigure[\label{QCDphases}]{%
\includegraphics[height=5.2cm,width=7.1cm]{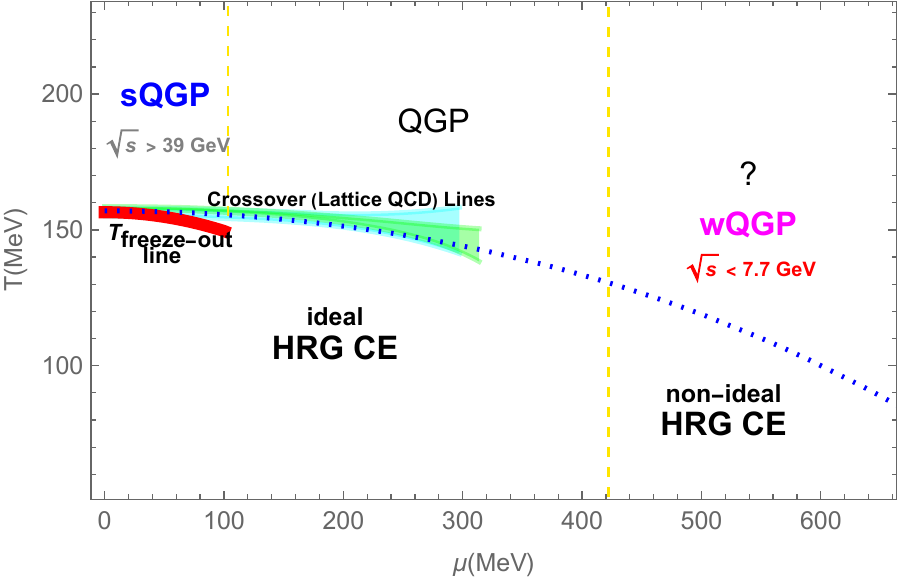}%
}
%\hfill
%\subfigure[\label{lambdas}]{%
%\includegraphics[height=4.84cm,width=7.1cm]{lambdas.pdf}%
%}
\caption{The QCD equation of state with various descriptions: The empirical freeze-out temperature $T_{\text{freeze-out}}(\mu)$, shown by the dotted \textcolor{blue}{blue} curve, follows (\ref{varyingT}). The shaded \textcolor{cyan}{cyan} and \textcolor{green}{green} curves represent the lattice pseudo-critical temperature of the QCD chiral crossover from \cite{HotQCD:2018pds} and \cite{Borsanyi:2020fev}, respectively. The \textcolor{red}{red} line corresponds to the lattice result of \cite{HotQCD:2018pds} after the replacement \eqref{replace} which reduces the regime of validity of the Taylor expansion method of lattice from $\mu_B\ll 300\,\text{MeV}$ to $\mu_B\ll 100\,\text{MeV}$.
}\label{lambdasQCDphases}
\end{figure}

The solid and dashed \textcolor{magenta}{magenta} curves, in Fig.\,\ref{CQCP}, show the net-proton number cumulant ratios defined by \eqref{firstC} for wQGP, in the nearly-cold $\mu \gg T$ limit. For the dashed \textcolor{magenta}{magenta} curves, we used the leading order (LO) $\alpha_s(\bar{\Lambda}) = 0$ limit of the free energy density \eqref{masslessp} in \eqref{firstC}. For the solid \textcolor{magenta}{magenta} curves, we used the full next-to-next-to-leading-order (NNLO) free energy density \eqref{masslessp} in \eqref{firstC}. Both dashed and solid \textcolor{magenta}{magenta} curves assume the varying freeze-out temperature and chemical potential (valid for $\mu\gg T_{freeze-out}$) given by 
\begin{align}\label{varyingT}
T_{\text{freeze-out}}(\sqrt{s}) &= 0.157\,\text{GeV} - 0.139\,\text{GeV}^{-1} \,\mu^2(\sqrt{s})\nonumber\\
&- 0.053\,\text{GeV}^{-3} \,\mu^4(\sqrt{s})\,,\\ 
\mu(\sqrt{s}) &= \frac{1.308\,\text{GeV}}{1 + 0.273\,\text{GeV}^{-1} \sqrt{s}}\,,\label{varyingmu}
\end{align}
taken from \cite{Cleymans:2005xv} by adjusting $T_{freeze-out}(\mu=0) = 0.157\,\text{GeV}$, see Fig.\,\ref{muTs}..
The solid \textcolor{magenta}{magenta} curves in Fig.\,\ref{CQCP} are our predictions for STAR Fixed-Target experiments at RHIC \cite{Meehan:2016qon}.

%\begin{widetext}
\begin{figure}
%[ht!]
\centering
\subfigure[\label{mus}]{%
\includegraphics[height=5cm,width=7.91cm]{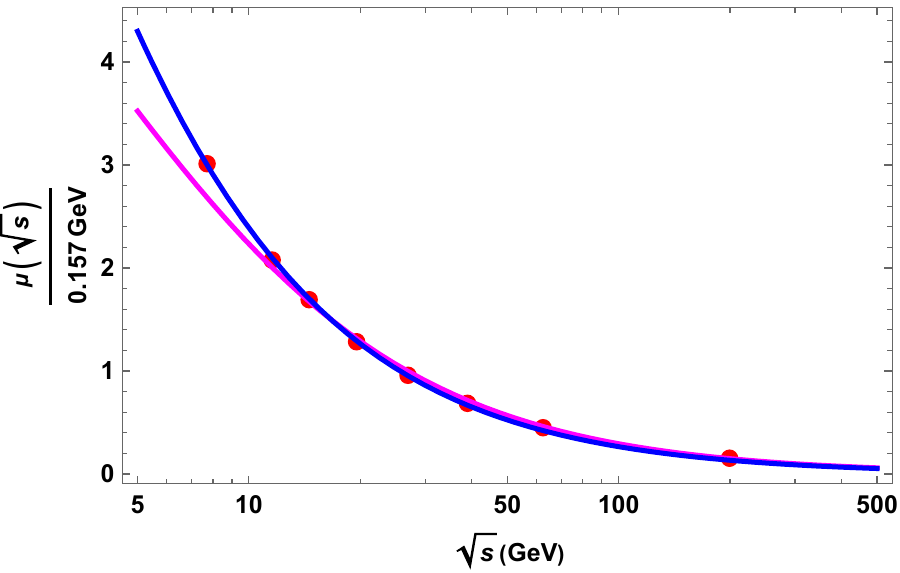}%
}
\hfill
\subfigure[\label{Ts}]{%
\includegraphics[height=5cm,width=7.91cm]{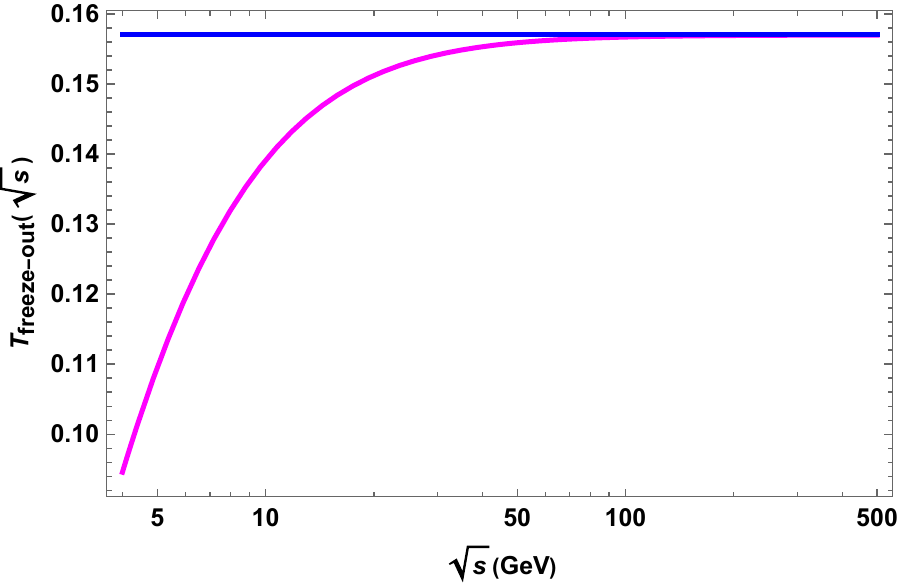}%
}
\caption{(a) The variation of the baryon chemical potential $\mu(\sqrt{s})$ using \eqref{mufixedTf} (\textcolor{blue}{Blue} curve which is a fit to the red data points), and using \eqref{varyingmu} (\textcolor{magenta}{Magenta} curve). (b) The variation of the freeze-out temperature $T_{freeze-out}(\sqrt{s})$ for fixed $T_{freeze-out}(\sqrt{s}) = 0.157\,\text{GeV}$ (\textcolor{blue}{Blue} curve), and using \eqref{varyingT} (\textcolor{magenta}{Magenta} curve). Note that the \textcolor{blue}{Blue} and \textcolor{magenta}{Magenta} curves are used for sQGP and wQGP cases, respectively.}
\label{muTs}
\end{figure}

\section{Summary and Outlook}
In this paper, without any adjustable free parameters fitted to lattice or experimental data, we have demonstrated that the \textit{dynamical soft-wall} AdS black hole model (sQGP) accurately describes the combined BES-I and BES-II experimental data \cite{Pandav:2024cpod} for net-proton number cumulant ratios for $\sqrt{s} \geq 39\,\text{GeV}$, as shown by the \textcolor{blue}{blue} curves in Fig.\,\ref{CQCP}.

In the intermediate energy range ($7.7\,\text{GeV} \leq \sqrt{s} \leq 39\,\text{GeV}$), the combined BES-I and BES-II experimental data \cite{Pandav:2024cpod} align with hydrodynamics \cite{Vovchenko:2021kxx} and hadron resonance gas \cite{Braun-Munzinger:2020jbk} models that include fixed or conserved baryon number in the canonical ensemble but do not feature a critical point, as indicated by the \textcolor{orange}{orange} and black curves in Fig.\,\ref{CQCP}.

Additionally, we have predicted that a nearly-cold ($\mu \gg T$) weakly coupled quark-gluon plasma (wQGP) can describe the data in the low energy regime ($\sqrt{s} < 7.7\,\text{GeV}$). This regime will be explored by the STAR Fixed-Target experiment at RHIC \cite{Meehan:2016qon}, as shown by the solid \textcolor{magenta}{magenta} curves in Fig.\,\ref{CQCP}. The wQGP result is consistent with the only available single data point in the low energy regime provided by the HADES collaboration \cite{HADES:2020wpc}.

Our findings provide a description of the QCD equation of state across different energy regimes. The \textit{dynamical soft-wall} model effectively captures the behavior at high energies, while the nearly-cold wQGP model extends our understanding to lower energies. Future experimental data from the STAR Fixed-Target experiments will further test these predictions and refine our understanding of the QCD phase diagram.

A summary of the various descriptions of the QCD equation of state is provided in Fig.\,\ref{lambdasQCDphases}, illustrating the different models and their applicable energy ranges.

\textit{ Acknowledgments}.---The author thanks Prithwish Tribedy for bringing the latest BES-II result to his attention. This work is supported by U.S. DOE Grant No. DE-FG02-04ER41302.

\bibliography{CriticalPoint}

\end{document}